\documentclass{article}
\usepackage{spconf}

\usepackage[latin1]{inputenc}
\usepackage{amssymb}
\usepackage{latexsym}
\usepackage{mathrsfs}
\usepackage{amsfonts}
\usepackage{amsbsy}
\usepackage{amsmath}
\usepackage{times}
\usepackage{epsfig,epsf,times,amssymb,amsmath}
 \ninept

\title{Reduced-rank Adaptive Constrained Constant Modulus Beamforming Algorithms based on Joint Iterative Optimization of Filters}
\name{\fontsize{11}{11}\selectfont\upshape Lei Wang and Rodrigo C. de Lamare \vspace{-0.75em}}
\address{\fontsize{10}{10}\selectfont\itshape
Communications Research Group, Department of Electronics \\
\fontsize{10}{10}\selectfont\itshape University of York, York YO10 5DD, UK
\\ \fontsize{9}{9}\selectfont\ttfamily\upshape
Email:\{lw517,rcdl500\}@ohm.york.ac.uk}

\begin{document}
\maketitle
\begin{abstract}
This paper proposes a reduced-rank scheme for adaptive beamforming based on the constrained joint iterative optimization of
filters. We employ this scheme to devise two novel reduced-rank adaptive algorithms according to the constant modulus (CM)
criterion with different constraints. The first devised algorithm is formulated as a constrained joint iterative optimization of
a projection matrix and a reduced-rank filter with respect to the CM criterion subject to a constraint on the array response.
The constrained constant modulus (CCM) expressions for the projection matrix and the reduced-rank weight vector are derived, and
a low-complexity adaptive algorithm is presented to jointly estimate them for implementation. The second proposed algorithm is
extended from the first one and implemented according to the CM criterion subject to a constraint on the array response and an
orthogonal constraint on the projection matrix. The Gram-Schmidt (GS) technique is employed to achieve this orthogonal
constraint and improve the performance. Simulation results are given to show superior performance of the proposed algorithms in
comparison with existing methods.


\end{abstract}

\section{Introduction}
Adaptive beamforming technology is of paramount importance in numerous signal processing applications such as radar, wireless
communications, and sonar \cite{Nemeth}, \cite{Anderson}. Among various beamforming techniques, the beamformers based on the
constrained minimum variance (CMV) criterion \cite{Jian} are prevalent and minimize the contribution of the total output power
while maintaining the gain along the direction of the signal of interest (SOI). Another alternative beamformer design is
performed according to the constrained constant modulus (CCM) \cite{Jian} criterion, which is a positive measure of the
beamformer output deviating from a constant modulus condition. Compared with the CMV, the CCM beamformers exhibit superior
performance in many severe scenarios (e.g., steering vector mismatch).

Many adaptive algorithms \cite{Jian} have been developed according to the CMV and CCM criteria for implementation. A simple and
popular one is the stochastic gradient (SG) method \cite{Frost}, \cite{Haykin}. However, the performance of the SG-based
algorithms is sensitive to the step size, the number of interferers and sensor elements, and the eigenvalue spread
\cite{Haykin}. For improving the performance, reduced-rank filtering has been introduced into beamforming in order to project
the received signal onto a lower dimension subspace and perform the filter optimization within this subspace. This technique
shows a fast convergence rate and improves tracking ability in situations where the number of sensor elements is large
\cite{Chen}. The Multi-stage Wiener filter (MSWF) \cite{Honig} and the auxiliary-vector filtering (AVF) \cite{Pados} are two
excellent approaches in this area. 
Employing these reduced-rank schemes, the
CMV beamformers reach improved performance but suffer from the heavy computational cost and instability. A joint iterative
optimization scheme \cite{Lamare} was presented recently with a simple adaptive implementation for reducing the complexity and
improving the tracking ability.

Considering the fact that the CCM-based beamformers outperform the CMV ones for constant modulus constellations, we propose two
adaptive reduced-rank algorithms according to the CCM criterion by employing a proposed reduced-rank scheme, which is based on
the constrained joint iterative optimization filters. The proposed algorithms are implemented according to the constant modulus
(CM) criterion with different constraints. The first one is formulated as a constrained joint iterative optimization of a
projection matrix and a reduced-rank filter with respect to the CM criterion subject to a constraint on the array response. The
projection matrix projects the received signal onto a lower dimension, which is then processed by the reduced-rank filter for
the array output. The CCM expressions for the projection matrix and the reduced-rank filter are derived, and a simple efficient
algorithm is presented to jointly estimate them for implementation. The second proposed algorithm is extended from the first one
and implemented according to the CM criterion subject to a constraint on the array response and an orthogonal constraint on the
projection matrix. We employ the Gram Schmidt (GS) technique \cite{Golub} to achieve this orthogonal constraint for the
projection matrix reformulation. The performance of the second algorithm outperforms the first one. Simulation results are given
to demonstrate the superior performance and stability achieved by the proposed algorithms versus the existing algorithms in
typical scenarios.

The remainder of this paper is organized as follows: we outline a system model for beamforming in Section 2. Based on this
model, the problem statement is presented. The proposed scheme, optimization and filter expressions are considered in Section 3.
Section 4 derives the proposed adaptive reduced-rank algorithms. The GS technique is briefly introduced in this part. Simulation
results are provided and discussed in Section 5, and conclusions are drawn in Section 6.

\section{System Model and Problem Statement}

\subsection{System Model}

Let us suppose that $q$ narrowband signals impinge on an uniform linear array (ULA) of $m$ ($m\geq q$) sensor elements. The
sources are assumed to be in the far field with directions of arrival (DOAs) $\theta_{0}$,\ldots,$\theta_{q-1}$. The $i$th
snapshot's vector of sensor array outputs $\boldsymbol x(i)\in\mathcal C^{m\times 1}$ can be modeled as
\begin{equation} \label{1}
\centering {\boldsymbol x}(i)={\boldsymbol {A}}({\boldsymbol {\theta}}){\boldsymbol s}(i)+{\boldsymbol n}(i),~~~ i=1,\ldots,N
\end{equation}
where $\boldsymbol{\theta}=[\theta_{0},\ldots,\theta_{q-1}]^{T}\in\mathcal{C}^{q \times 1}$ is the signal DOAs, ${\boldsymbol
A}({\boldsymbol {\theta}})=[{\boldsymbol a}(\theta_{0}),\ldots,{\boldsymbol a}(\theta_{q-1})]\in\mathcal{C}^{m \times q}$
comprises the signal direction vectors ${\boldsymbol a}(\theta_{k})=[1,e^{-2\pi
j\frac{d}{\lambda_{c}}cos{\theta_{k}}},\ldots$,\\$e^{-2\pi j(m-1)\frac{d}{\lambda_{c}}cos{\theta_{k}}}]^{T}\in\mathcal{C}^{m
\times 1}, (k=0,\ldots,q-1)$, where $\lambda_{c}$ is the wavelength and $d$ is the inter-element distance of the ULA
($d=\lambda_c/2$ in general), and to avoid mathematical ambiguities, the direction vectors $\boldsymbol a(\theta_{k})$ are
considered to be linearly independents. ${\boldsymbol s}(i)\in \mathcal{C}^{q\times 1}$ is the source data, ${\boldsymbol
n}(i)\in\mathcal{C}^{m\times 1}$ is temporarily white sensor noise, which is assumed to be a zero-mean spatially and Gaussian
process, $N$ is the observation size of snapshots, and $(\cdot)^{T}$\ stands for transpose. The output of a narrowband
beamformer is given by
\begin{equation} \label{2}
\centering y(i)={\boldsymbol w}^H(i) {\boldsymbol x}(i)
\end{equation}
where ${\boldsymbol w}(i)=[w_{1}(i),\ldots,w_{m}(i)]^{T}\in\mathcal{C}^{m\times 1}$ is the complex weight vector, and
$(\cdot)^{H}$ stands for Hermitian transpose.

\subsection{Problem Statement}

Let us consider the full-rank CCM optimization filter for beamforming, which can be computed by solving the following
optimization problem
\begin{equation}\label{3}
\begin{split}
&{\boldsymbol w}_{\textrm{opt}}=\arg\min_{\boldsymbol w} E\big\{\big[|y(i)|^p-R_{p}\big]^2\big\},~i=1,\ldots,N\\
&\textrm{subject~to}~~{\boldsymbol w}^{H}(i){\boldsymbol a}(\theta_{0})=1.
\end{split}
\end{equation}
where the constant $R_{p}$ is suitably chosen to guarantee that the weight solution is close to the global minimum and the
constraint is set to ensure a closed-form solution. The quantity $\theta_0$ is the direction of the SOI, $\boldsymbol
a(\theta_{0})$ denotes the normalized steering vector of the desired signal, and in general, $p=2$ is selected to consider the
optimization as the expected deviation of the squared modulus of the array output to a constant, say $R_{p}=1$. The CCM
beamformer minimizes the contribution of undesired interference while maintaining the gain along the look direction to be
constant. Using the method of Lagrange multipliers to solve the optimization problem in (\ref{3}), the weight expression is
\begin{equation}\label{4}
\boldsymbol w=\frac{\boldsymbol R^{-1}\boldsymbol a(\theta_{0})}{\boldsymbol a^{H}(\theta_{0})\boldsymbol R^{-1}\boldsymbol
a(\theta_{0})}
\end{equation}
where $\boldsymbol R=E[2(|y(i)|^{2}-1)\boldsymbol x(i)\boldsymbol x^{H}(i)]\in\mathcal C^{m\times m}$ is the expected cross
correlation matrix between $\boldsymbol x(i)$ and $y(i)$. The complexity can be high due to the existence of the covariance
matrix inverse. In practice, $\boldsymbol R$ is not available but has to be estimated, which may result in the poor convergence
and tracking ability when $m$ is large. Note that $\boldsymbol R$ depends on $y(i)$, which is a function of current $\boldsymbol
w(i)$. By initializing $\boldsymbol w(i)$ and estimating a prior $y(i)$, we can estimate $\boldsymbol R$ and get the weight
solution for each snapshot.

\section{Proposed Reduced-rank Scheme and CCM Filters Design}
In this section, we employ a reduced-rank scheme to introduce two optimization problems according to the CM criteria subject to
different constraints. The reduced-rank scheme is based on a constrained joint iterative optimization of a projection matrix and
a reduced-rank filter. The CCM expressions of the projection matrix and the reduced-rank weight vector are derived.

\subsection{Proposed Reduced-Rank Scheme and Optimization Problems}
Consider a projection matrix $\boldsymbol T_{r}(i)=[\boldsymbol t_1(i), \boldsymbol t_2(i), \ldots,$ $\boldsymbol
t_r(i)]\in\mathcal C^{m\times r}$, which is responsible for the dimensionality reduction, to project the $m\times 1$ input
vector $\boldsymbol x(i)$ onto a lower dimension, yielding
\begin{equation}\label{5}
\bar{\boldsymbol x}(i)=\boldsymbol T_{r}(i)^{H}\boldsymbol x(i)
\end{equation}
where $\boldsymbol t_l(i)=[t_{1,l}(i), \ldots, t_{m,l}(i)]^T\in\mathcal C^{m\times 1},~l=1, \ldots, r$, makes up the projection
matrix $\boldsymbol T_r(i)$, $\bar{\boldsymbol x}(i)\in\mathcal C^{r\times 1}$ is the projected input vector, and in what
follows, all $r$-dimensional quantities are denoted by an over bar. Here, $r<m$ is the rank and, as we will see, impacts the
output performance. An adaptive reduced-rank filter represented by $\bar{\boldsymbol
w}(i)=[\bar{w}_{1}(i),\ldots,\bar{w}_{r}(i)]^{T}\in\mathcal{C}^{r\times 1}$ is followed to process the projected data for
estimating the output
\begin{equation}\label{6}
y(i)=\bar{\boldsymbol w}^H(i)\boldsymbol T_r^{H}(i)\boldsymbol x(i)
\end{equation}

\begin{figure}[!t]
\begin{center}
\includegraphics[angle=0,
width=0.5\textwidth,height=0.12\textheight]{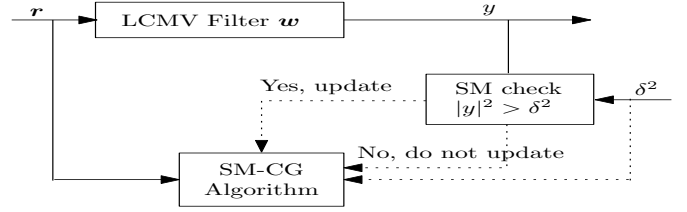}
\end{center} \caption{\label{fig:model1} Proposed reduced-rank beamforming scheme.}
\end{figure}

From (\ref{6}), the array output $y(i)$ depends on the projection matrix $\boldsymbol T_r(i)$ and the reduced-rank weight vector
$\bar{\boldsymbol w}(i)$, as shown in Fig. \ref{fig:model1}. It is necessary to jointly optimize $\boldsymbol T_r(i)$ and
$\bar{\boldsymbol w}(i)$ to estimate $y(i)$. We consider two optimization problems, which are problem i)
\begin{equation}\label{7}
\begin{split}
&[\boldsymbol T_{r,\textrm{opt}}, \bar{\boldsymbol w}_{\textrm{opt}}]=\arg\min_{\boldsymbol T_r,\bar{\boldsymbol
w}}E\big\{\big[|y(i)|^2-1\big]^2\big\},~i=1, \ldots, N\\
&\textrm{subject to}~\bar{\boldsymbol w}^H(i)\boldsymbol T_r(i)\boldsymbol a(\theta_0)=1.
\end{split}
\end{equation}

and problem ii)
\begin{equation}\label{8}
\begin{split}
&[\boldsymbol T_{r,\textrm{opt}}, \bar{\boldsymbol w}_{\textrm{opt}}]=\arg\min_{\boldsymbol T_r,\bar{\boldsymbol
w}}E\big\{\big[|y(i)|^2-1\big]^2\big\},~i=1, \ldots, N\\
&\textrm{subject to}~\bar{\boldsymbol w}^H(i)\boldsymbol T_r(i)\boldsymbol a(\theta_0)=1~\textrm{and}~\boldsymbol
T_r(i)^H\boldsymbol T_r(i)=\boldsymbol I.
\end{split}
\end{equation}

Compared with problem i), problem ii) includes one orthogonal constraint on the projection matrix, which is to reformulate
$\boldsymbol T_r(i)$ for improving the performance. In the following part, we will derive the CCM expressions of $\boldsymbol
T_r(i)$ and $\bar{\boldsymbol w}(i)$ with respect to problem i). The proposed adaptive algorithm for the implementation of
(\ref{7}) and the extended algorithm for problem ii) will represent in Section 4.

\subsection{Design of CCM Filters}
The constraint in (\ref{7}) can be incorporated by the method of Lagrange multipliers \cite{Haykin} in the form
\begin{equation}\label{9}
\mathcal J=E\big\{\big[|y(i)|^{2}-1\big]^{2}\big\}+\lambda\big[\bar{\boldsymbol w}^H(i)\boldsymbol T_r^{H}(i){\boldsymbol
a}(\theta_0)-1\big]
\end{equation}
where $\lambda$ is a scalar Lagrange multiplier. Substituting (\ref{6}) into (\ref{9}), fixing $\bar{\boldsymbol w}(i)$, taking
the gradient of (\ref{9}) with respect to $\boldsymbol T_r(i)$, and setting it equals to a null matrix, yields
\begin{equation}\label{10}
\nabla\mathcal J_{T_r}=\boldsymbol R\boldsymbol T_r(i)\bar{\boldsymbol R}_{w}+\lambda_{T_r}\boldsymbol
a(\theta_{0})\bar{\boldsymbol w}^H(i)
\end{equation}
where $e(i)=|y(i)|^{2}-1$, $\boldsymbol R=E[2e(i)\boldsymbol x(i)\boldsymbol x^H(i)]$ is the expected cross correlation matrix,
and $\bar{\boldsymbol R}_w=E[\bar{\boldsymbol w}(i)\bar{\boldsymbol w}^H(i)]$ is the expected reduced-rank weight matrix. Both
$\boldsymbol R$ and $\bar{\boldsymbol R_w}$ need to be estimated by sample-averaging in practice. Note that $\boldsymbol R$
depends on $y(i)$, which is a function of $\boldsymbol T_r(i)$ and $\bar{\boldsymbol w}(i)$. By initializing $\boldsymbol
T_r(i)$ and $\bar{\boldsymbol w}(i)$ and using a prior $y(i)$, we can estimate $\boldsymbol R$.

Rearranging the second equation of (\ref{10}) to represent $\boldsymbol T_r(i)$, which is then substituted into the constraint
in (\ref{7}) for solving the Lagrange multiplier $\lambda_{{T}_r}$, we get the result for the projection matrix
\begin{equation}\label{11}
\boldsymbol T_r(i)=\frac{\boldsymbol R^{-1}\boldsymbol a(\theta_{0})\bar{\boldsymbol w}^H(i)\bar{\boldsymbol
R}_w^{-1}}{\bar{\boldsymbol w}^H(i)\bar{\boldsymbol R}_w^{-1}\bar{\boldsymbol w}(i)\boldsymbol a^H(\theta_{0})\boldsymbol
R^{-1}\boldsymbol a(\theta_{0})}
\end{equation}

On the other hand, fixing $\boldsymbol T_{r}(i)$, taking the gradient of (\ref{9}) with respect to $\bar{\boldsymbol w}(i)$, and
setting it equal to a null vector, we have
\begin{equation}\label{12}
\nabla\mathcal J_{\bar{w}}=\bar{\boldsymbol R}\bar{\boldsymbol w}(i)+\lambda_{\bar{w}}\boldsymbol T_r^H(i)\boldsymbol
a(\theta_{0})
\end{equation}
where $\bar{\boldsymbol R}=E[2e(i)\bar{\boldsymbol x}(i)\bar{\boldsymbol x}^H(i)]\in\mathcal C^{r\times r}$ is the expected
reduced-rank cross correlation matrix, which is estimated by sample-averaging. Following the same procedures for $\boldsymbol
T_r(i)$, the result for the reduced-rank weight solution can be expressed as
\begin{equation}\label{13}
\bar{\boldsymbol w}(i)=\big[\bar{\boldsymbol a}^H(\theta_{0})\bar{\boldsymbol R}^{-1}\bar{\boldsymbol
a}(\theta_{0})\big]^{-1}\bar{\boldsymbol R}^{-1}\bar{\boldsymbol a}(\theta_{0})
\end{equation}
where $\bar{\boldsymbol a}(\theta_{0})=\boldsymbol T_r^H(i)\boldsymbol a(\theta_{0})\in\mathcal C^{r\times 1}$ is the projected
steering vector of the SOI.

The update equations (\ref{11}) for the projection matrix and (\ref{13}) for the reduced-rank weight vector depend on each other
and so are not closed-form solutions. It is necessary to iterate $\boldsymbol T_r$ and $\bar{\boldsymbol w}$ with initial values
for implementation. Therefore, the initialization is not only for obtaining a prior $y$ but starting the iteration of the
proposed scheme. The projection matrix creates a connection between the full-rank input vectors and the reduced-rank ones,
whereas the reduced-rank filter recovers the transmitted signal. They are jointly updated to solve the CCM optimization problem
i), i.e., the so-called ``\textit{joint iterative optimization}" (JIO).

\section{Development of Adaptive Algorithms}
\subsection{Proposed Adaptive SG Algorithm for Problem i)}
We describe a simple adaptive algorithm for implementation of the proposed reduced-rank scheme based on the optimization problem
i). Fixing $\bar{\boldsymbol w}(i)$ and $\boldsymbol T_r(i)$, respectively, taking the instantaneous gradient of (\ref{7}) with
respect to $\boldsymbol T_r(i)$ and $\bar{\boldsymbol w}(i)$, and setting them equal to null, we obtain
\begin{equation}\label{14}
\nabla\mathcal J_{p,T_r}=2e(i)y^{*}(i)\boldsymbol x(i)\bar{\boldsymbol w}^H(i)+\lambda_{p,T_r}\boldsymbol
a(\theta_{0})\bar{\boldsymbol w}^H(i)
\end{equation}
\begin{equation}\label{15}
\nabla\mathcal J_{p,\bar{w}}=2e(i)y^{*}(i)\boldsymbol T_r^H(i)\boldsymbol x(i)+\lambda_{p,\bar{w}}\boldsymbol
T_r^H(i)\boldsymbol a(\theta_{0})
\end{equation}
where the subscript ``p" means the proposed and $(\cdot)^{*}$ denotes complex conjugate.

Following the gradient rules $\boldsymbol T_r(i+1)=\boldsymbol T_r(i)-\mu_{T_r}\nabla\mathcal J_{p,T_r}$ and $\bar{\boldsymbol
w}(i+1)=\bar{\boldsymbol w}(i)-\mu_{\bar{w}}\nabla\mathcal J_{p,\bar{w}}$, substituting (\ref{14}) and (\ref{15}) into them,
respectively, and solving the Lagrange multipliers $\lambda_{p,T_r}$ and $\lambda_{p,\bar{w}}$ by employing the constraint in
(\ref{7}), we obtain the iterative solutions in the form
\begin{equation}\label{16}
\begin{split}
\boldsymbol T_r(i+1)=\boldsymbol T_r(i)-&\mu_{T_r}e(i)y^*(i)\big[\boldsymbol x(i)\bar{\boldsymbol w}^H(i)\\
&-\boldsymbol a(\theta_{0})\bar{\boldsymbol w}^H(i)\boldsymbol a^H(i)\boldsymbol x(i)\big]
\end{split}
\end{equation}
\begin{equation}\label{17}
\bar{\boldsymbol w}(i+1)=\bar{\boldsymbol w}(i)-\mu_{\bar{w}}e(i)y^*(i)\big[\boldsymbol I-\frac{\bar{\boldsymbol
a}(\theta_{0})\bar{\boldsymbol a}^H(\theta_{0})}{\bar{\boldsymbol a}^H(\theta_{0})\bar{\boldsymbol
a}(\theta_{0})}\big]\bar{\boldsymbol x}(i)
\end{equation}
where $\mu_{T_r}$ and $\mu_{\bar{w}}$ are the corresponding step sizes, which are small positive values. The projection matrix
$\boldsymbol T_r(i)$ and the reduced-rank weight vector $\bar{\boldsymbol w}(i)$ are jointly updated. The output $y(i)$ at time
instant $i$ can be estimated after each joint optimization procedure with respect to the CCM criterion. We denominate this
proposed algorithm ((\ref{16}) and (\ref{17})) as JIO-CCM.

\subsection{Extended Algorithm for Problem ii)}
Now, we consider the optimization problem ii). As explained before, the constraint is added to orthogonalize a set of vectors
$\boldsymbol t_l(i)$ for the performance improvement. We employ the Gram-Schmidt (GS) technique \cite{Golub} to realize this
constraint. Specifically, the adaptive SG algorithm in (\ref{16}) is implemented to obtain $\boldsymbol T_r(i+1)$. Then, the GS
process is performed to reformulate the projection matrix, which is \cite{Golub}
\begin{equation}\label{18}
\boldsymbol t_{l,\textrm{ort}}(i)=\boldsymbol
t_{l}(i)-\sum_{j=1}^{l-1}\textrm{proj}_{\boldsymbol
t_{j,\textrm{ort}}(i)}\boldsymbol t_l(i)
\end{equation}
where $\boldsymbol t_{l,\textrm{ort}}(i)$ is the normalized orthogonal vector after the GS process and
$\textrm{proj}_{\boldsymbol t_{j,\textrm{ort}}(i)}\boldsymbol t_l(i)={\boldsymbol t_{j,\textrm{ort}}^H(i)}\boldsymbol
t_l(i)\frac{{\boldsymbol t_{j,\textrm{ort}}}(i)}{\boldsymbol t_{j,\textrm{ort}}^H(i)\boldsymbol t_{j,\textrm{ort}}(i)}$ is a
projection operator.

The reformulated projection matrix $\boldsymbol T_{r,\textrm{ort}}(i)$ is constructed when we obtain a set of orthogonal
$\boldsymbol t_{l, \textrm{ort}}(i),~l=1, \ldots, r$. By employing $\boldsymbol T_{r,\textrm{ort}}(i)$ to get $\bar{\boldsymbol
x}(i)$, $\bar{\boldsymbol a}(\theta_0)$, and jointly update with $\bar{\boldsymbol w}(i+1)$ in (\ref{17}), the performance can
be further improved. Simulation results will be given for showing this result. We denominate this GS version algorithm as
JIO-CCM-GS, which is performed by computing (\ref{16}), (\ref{18}), and (\ref{17}).

\subsection{Computational Complexity}

The computational complexity with respect to the existing and proposed algorithms is evaluated according to additions and
multiplications. The complexity comparison is listed in Table \ref{tab: Computational complexity}. The complexity of the
proposed JIO-CCM and JIO-CCM-GS algorithms increases with the multiplication of $rm$, specifically $m$ since the rank $r$ is
selected around a small range that is much less than $m$ in large arrays' conditions without performance degradation. This fact
will be shown in the simulation. This complexity is about $r$ times higher than the full-rank algorithms \cite{Frost},
\cite{Haykin}, slightly higher than the recent JIO-CMV based algorithm \cite{Lamare}, but much lower than the MSWF-based
\cite{Honig}, \cite{Lamare2}, and AVF \cite{Pados} methods.
\begin{table}
\centering \caption{\normalsize Computational complexity} \footnotesize \label{tab: Computational complexity}
\begin{tabular}{l c c}
\hline \\
Algorithm & Additions                   & Multiplications\\
\hline   \\
Full-Rank-CMV    & $3m-1$                        & $4m+1$ \\
Full-Rank-CCM    & $3m$                  & $4m+3$ \\
MSWF-CMV         & $rm^2+rm+m$             & $rm^{2}+m^2+2rm$ \\
                 & $+2r-2$                            & $+5r+2$\\
MSWF-CCM        & $rm^2+rm+m$             & $rm^2+m^2+2rm$ \\
                & $+2r-1$                 & $+5r+4$\\
AVF           & $r(4m^2+m-2)$   & $r(5m^2+3m)$\\
              & $+5m^2-m-1$     & $+8m^2+2m$\\
JIO-CMV        & $4rm+m+2r-3$            & $4rm+m+7r+3$\\
JIO-CMV-GS     & $7rm-m-1$               & $7rm-2m+8r+2$\\
JIO-CCM        & $4rm+m+2r-2$             & $4rm+m+7r+6$\\
JIO-CCM-GS     & $7rm-m$                  & $7rm-2m+8r+5$\\
\hline
\end{tabular}
\end{table}

\section{Simulations}

Simulations are performed by an ULA containing $m=32$ sensor
elements with half-wavelength interelement spacing. We compare the
proposed JIO-CCM and JIO-CCM-GS algorithms with the full-rank
\cite{Haykin}, MSWF \cite{Honig}, \cite{Lamare2}, and AVF
\cite{Pados} methods and in each method, the CMV and CCM criterions
are considered with SG algorithms for implementation. A total of
$K=1000$ runs are used to get the curves. In all experiments, the
BPSK source power (including the desired user and interferers) is
$\sigma_{s}^{2}=\sigma_{i}^2=1$ and the input SNR $=10$ dB with
spatially and temporally white Gaussian noise.

\begin{figure}[!htb]
\begin{center}
\def\epsfsize#1#2{1.0\columnwidth}
\epsfbox{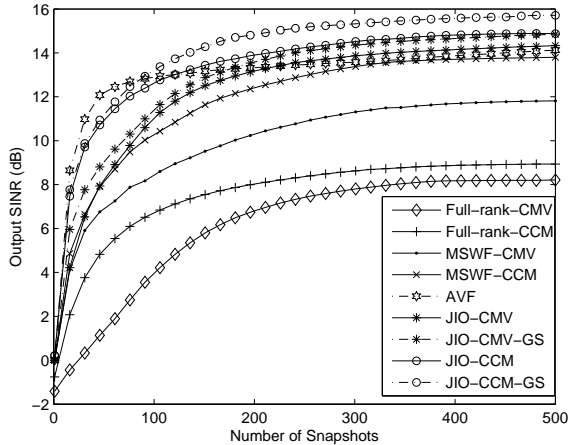} \caption{Output SINR versus the number of
snapshots with $m=32$, $q=7$, SNR$=10$ dB, $\mu_{T_r}=0.002$,
$\mu_{\bar{w}}=0.001$, $\mu_{T_r,\textrm{gs}}=0.003$,
$\mu_{\bar{w},\textrm{gs}}=0.0007$.}
\label{fig:cmv_ccm_sg_gram_final}
\end{center}
\end{figure}

In Fig. \ref{fig:cmv_ccm_sg_gram_final}, we consider the presence of $q=7$ users (one desired) in the system. The projection
matrix and the reduced-rank weight vector are initialized with $\boldsymbol T_r(0)=[\boldsymbol I_r^T~\boldsymbol
0_{r\times(m-r)}^T]$ and $\bar{\boldsymbol w}(0)=\big(\boldsymbol T_{r}^{H}(0)\boldsymbol a(\theta_{0})\big)/\big(\|\boldsymbol
T_{r}^{H}(0)\boldsymbol a(\theta_{0})\|^{2}\big)$ to ensure the constraint in (\ref{7}). The rank is $r=r_{\textrm gs}=5$  for
the proposed JIO-CCM and JIO-CCM-GS algorithms. The expected matrix $\boldsymbol R$ used in the MSWF and AVF is estimated by
sample-averaging. Fig. \ref{fig:cmv_ccm_sg_gram_final} shows that all output SINR curves increase to steady-state as increase of
the snapshots. The joint optimization based algorithms have superior performance as compared with the full-rank, MSWF, and AVF
methods. Their GS version algorithms enjoy further developed performance comparing with corresponding JIO-CMV and JIO-CCM
methods. The proposed JIO-CCM and JIO-CCM-GS algorithms outperforms the existing methods in the output performance. Checking the
convergence, the proposed algorithms are slightly slower than the AVF, which is least squares (LS)-based, and much faster than
the other methods.

\begin{figure}[!htb]
\begin{center}
\def\epsfsize#1#2{1.0\columnwidth}
\epsfbox{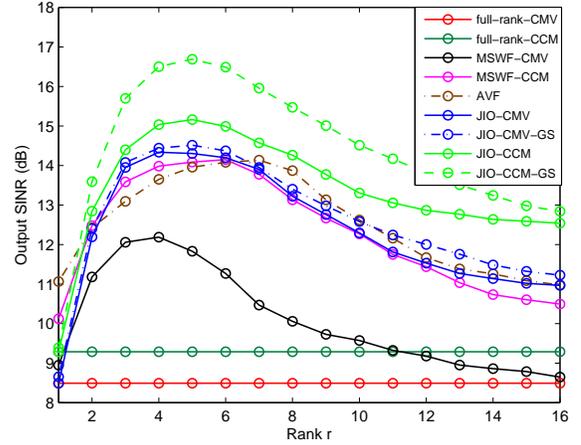} \caption{Output SINR versus rank ($r$) with
$m=32$, $q=7$, SNR$=10$ dB, $N=500$, $\mu_{T_r}=0.002$,
$\mu_{\bar{w}}=0.001$, $\mu_{T_r,\textrm{gs}}=0.003$,
$\mu_{\bar{w},\textrm{gs}}=0.0007$.}
\label{fig:cmv_ccm_gram_rank_final}
\end{center}
\end{figure}

In Fig. \ref{fig:cmv_ccm_gram_rank_final}, we keep the same scenario
as that in Fig. \ref{fig:cmv_ccm_sg_gram_final} and check the rank
selection for the existing and proposed algorithms. The number of
snapshots is fixed to $N=500$. The optimum choices for the proposed
algorithms are $r=r_{\textrm{gs}}=5$, which are comparatively lower
than most existing algorithms, but reach superior performance. We
also checked the change of these values for different array sizes
and data records, and verified that they are nearly invariant, which
saves computation cost.

\begin{figure}[!htb]
\begin{center}
\def\epsfsize#1#2{1.0\columnwidth}
\epsfbox{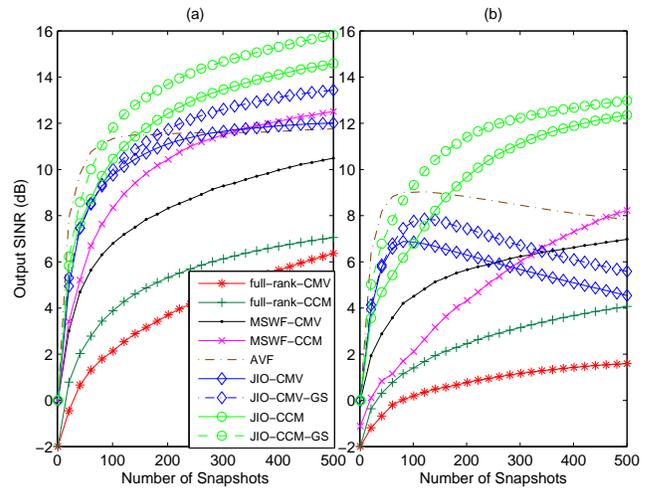} \caption{Output SINR versus the number of
snapshots with $m=32$, $q=10$, SNR$=10$ dB, $\mu_{T_r}=0.002$,
$\mu_{\bar{w}}=0.001$, $\mu_{T_r,\textrm{gs}}=0.003$,
$\mu_{\bar{w},\textrm{gs}}=0.0007$ for (a) ideal steering vector
condition; (b) steering vector mismatch $2^o$.}
\label{fig:cmv_ccm_sve_gram_final}
\end{center}
\end{figure}

Finally, the mismatch (steering vector error) condition is analyzed in Fig. \ref{fig:cmv_ccm_sve_gram_final}. Here, the number
of users is $q=10$, including one desired user. In Fig. \ref{fig:cmv_ccm_sve_gram_final}(a), the exact DOA of the SOI is used in
the algorithms. The output performance of the proposed algorithms is better than those of the existing algorithms, and the
convergence is a little slower than that of the AVF algorithm, but higher than the others. In Fig.
\ref{fig:cmv_ccm_sve_gram_final}(b), we set the DOA of the SOI estimated by the receiver to be $2^{\textit o}$ away from the
actual direction. This indicates that the mismatch problem induces performance degradation to all the analyzed algorithms. The
CCM-based methods are more robust to this scenario than the CMV-based ones. The proposed algorithms still retain outstanding
performance compared with other techniques.

\section{Concluding Remarks}

We proposed a reduced-rank scheme based on the joint iterative optimization filters for beamforming and devised two adaptive
reduced-rank algorithms according to the CCM criteria, namely, JIO-CCM and JIO-CCM-GS. They are implemented by employing a
low-complexity adaptive algorithm to jointly update the projection matrices and reduced-rank filters. The JIO-CCM-GS algorithm,
by reformulating the projection matrix, achieves faster convergence and better performance than the JIO-CCM. The GS technique is
employed to realize the reformulation. The devised algorithms, compared with the existing methods, show preferable performance
and fast convergence in the studied scenarios.

\end{document}